\documentclass{elsart}
 \usepackage{latexsym}
 \usepackage{epsfig}
\usepackage{amssymb}

\newcommand{\be}{\begin{equation}}
\newcommand{\ee}{\end{equation}}
\newcommand{\dst}{\displaystyle}
\newcommand{\fr}[2]{\frac{{\dst #1}}{{\dst #2}}}

\date{March 15, 2009}

\begin{document}
\begin{frontmatter}

\title{Beam-size effect and particle losses \\at Super$B$ factory (Italy)}

\author{G.L.~Kotkin, V.G.~Serbo\thanksref{ead}}
 \thanks[ead]{Corresponding author. E-mail: serbo@math.nsc.ru}
%\author{V.G.~Serbo}%\corauthref{cor}}
%\corauth[cor]{Corresponding author. E-mail:
%\ead{serbo@math.nsc.ru}

\address{Novosibirsk State University, 630090 Novosibirsk, Russia}
%\begin{document}
%\maketitle

\begin{abstract}
In the colliders, the macroscopically large impact parameters give
a substantial contribution to the standard cross section of the
$e^+ e^- \rightarrow e^+ e^- \gamma$ process. These impact
parameters may be much larger than the transverse sizes of the
colliding bunches. It means that the standard cross section of
this process has to be substantially modified. In the present
paper such a beam-size effect is calculated for bremsstrahlung at
Super$B$ factory developed in Italy. We find out that this effect
reduces beam losses due to bremsstrahlung by about 40\%.
\end{abstract}
\end{frontmatter}

%%%%%%%%%%%%%%%%%%%%%%%%%%%%%%%%%%%%%%%%%%%%%%%
\section{Introduction: beam-size or MD-effect}
%%%%%%%%%%%%%%%%%%%%%%%%%%%%%%%%%%%%%%%%%%%%%%%%%

The so called beam-size or MD-effect is a phenomenon discovered in
experiments \cite{Blinov82} at the MD-1 detector (the VEPP-4
accelerator with $e^+e^-$ colliding beams , Novosibirsk 1981). It
was found that for ordinary bremsstrahlung, macroscopically large
impact parameters should be taken into consideration. These impact
parameters may be much larger than the transverse sizes of the
interacting particle bunches. In that case, the standard
calculations, which do not take into account this fact, will give
incorrect results. The detailed description of the MD-effect can
be found in review \cite{KSS}.

In the present paper we calculate the MD-effect and its influence
on the beam particle losses at the Super$B$ factory developed in
Italy~\cite{CDR}. We find out that this effect reduces beam losses
due to bremsstrahlung by about 40\%. For the reader convenience,
we repeat briefly historical introduction from our
paper~\cite{KS-2005}.

In $1980$--$1981$ a dedicated study of the process $e^+ e^-
\rightarrow e^+ e^- \gamma$ has been performed at the collider
VEPP-4 in Novosibirsk using the detector MD-1 for an energy of the
electron and positron beams $E_e=E_p = 1.8$ GeV and in a wide
interval of the photon energy $E_\gamma$ from $0.5$ MeV to
$E_\gamma \approx E_e$. It was observed \cite{Blinov82} that the
number of measured photons was smaller than expected. The
deviation from the standard calculation reached $30 \%$ in the
region of small photon energies and vanished for large energies of
the photons. A qualitative explanation of the effect was given by
Yu.A.~Tikhonov \cite{Tikhonov82}, who pointed out that those
impact parameters $\varrho$, which give an essential contribution
to the standard cross section, reach values of $\varrho_m \sim 5$
cm whereas the transverse size of the bunch is $\sigma_\perp \sim
10^{-3}$ cm. The limitation of the impact parameters to values
$\varrho \lesssim \sigma_\perp$ is just the reason for the
decreasing number of observed photons.

The first calculations of this effect have been performed in Refs.
\cite{BKS} and \cite{BD} using different versions of
quasi--classical calculations in the region of large impact
parameters. Later on, the effect of limited impact parameters was
taken into account using the single bremsstrahlung reaction for
measuring the luminosity at the VEPP--$4$ collider~\cite{Blinov88}
and at the LEP-I collider~\cite{Bini94}.

A general scheme  to calculate the finite beam size effect had
been developed in paper~\cite{KPS85a} starting from the quantum
description of collisions as an interaction of wave packets that
form bunches. It has also been shown that similar effects have to
be expected for several other reactions such as bremsstrahlung for
colliding $ep$--beams~\cite{KPS85b}, \cite{KPSS88}, $e^+e^-$--
pair production in $e^\pm e$ and $\gamma e$
collisions~\cite{KPS85a}.

In 1995 the MD-effect was experimentally  observed at the
electron-proton collider HERA~\cite{Piot95} at the level predicted
in~\cite{KPSS88}.

It was realized in last years that  the MD-effect in
bremsstrahlung plays an important role in the beam lifetime
problem. At storage rings TRISTAN and LEP-I, the process of single
bremsstrahlung was the dominant mechanism for the particle losses
in beams. If electron loses more than $1\;\%$ of its energy, it
leaves the beam. Since the MD-effect considerably reduced the
effective cross section of this process, the calculated beam
lifetime in these storage rings was larger by about $25 \; \%$ for
TRISTAN~\cite{Funakoshi} and by about $40 \; \%$ for
LEP-I~\cite{Burkhardt-1993} (in accordance with the experimental
data) then without taken into account the MD-effect. According to
our calculations~\cite{KS-2005}, at B-factories PEP-II and KEKB
the MD effect reduces beam losses due to brems\-strah\-lung by
about 20~\%.

In next Section we give the qualitative description of the
MD-effect. In Sec. 3 we present our results for Super$B$
factory~\cite{CDR}. In the last Section we compare our results
with those presented in Sec. 3.6.2 of the paper~\cite{CDR}. Though
we find a good agreement we argue that this agreement is just a
random coincidence because the basic ideas and formulas for these
two results are quite different. We present strong arguments in
favor of our approach.

Below we use the following notations: $E_e$ and $E_p$ are the
energies of the electron and positron, $N_e$ and $N_p$ are the
numbers of electrons and positrons in the bunches, $\sigma_H$ and
$\sigma_V$ are the horizontal and vertical transverse sizes of the
bunch, $r_e=e^2/(m_e c^2)$ is the classical electron radius,
$\gamma_e=E_e/(m_ec^2)$, $\gamma_p=E_p/(m_pc^2)$ and $\alpha
\approx 1/137$.

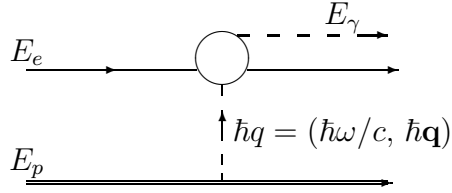
\begin{figure}[!t]
  \centering
  \setlength{\unitlength}{1cm}
\unitlength=2.0mm \special{em:linewidth 0.4pt}
\linethickness{0.4pt}
\begin{picture}(26.00,15.00)

\put(1.00,1.80){\line(1,0){24.00}}
\put(1.00,1.60){\line(1,0){24.00}}
\put(25.2,1.70){\vector(1,0){0.10}}

\put(1.00,9.20){\vector(1,0){6.00}}
\put(14.00,10.00){\circle{3.40}} \put(7.00,9.2){\line(1,0){5.3}}
\put(15.70,9.20){\vector(1,0){10.00}}
\put(15.0,11.5){\line(1,0){0.8}} \put(17,11.5){\line(1,0){0.8}}
\put(19,11.5){\line(1,0){0.8}} \put(21,11.5){\line(1,0){0.8}}
\put(23,11.5){\vector(1,0){2.00}}

\put(14.00,4.50){\vector(0,1){2.00}}
\put(14.00,2.90){\line(0,1){0.71}}
\put(14.00,1.80){\line(0,1){0.51}}
\put(14.00,7.50){\line(0,1){0.7}}

\put(1.00,10.4){\makebox(0,0)[cc]{$E_e$}}
\put(1.00,3.00){\makebox(0,0)[cc]{$E_p$}}
\put(22.00,12.70){\makebox(0,0)[cc]{$E_\gamma$}}
\put(22.00,5.00){\makebox(0,0)[cc]{$\hbar q=(\hbar\omega/c,\,
\hbar{\bf q})$}}

\end{picture}
    \caption{Block diagram of radiation by the electron.}
 \label{fig:1}
  \end{figure}

  \begin{figure}[!t]
  \centering
\unitlength=2.00mm \special{em:linewidth 0.4pt}
\linethickness{0.4pt}
\begin{picture}(47.00,15.00)
\put(2.00,9.20){\vector(1,0){2.00}}
\put(10.00,10.00){\circle{3.40}} \put(3.00,9.20){\line(1,0){5.30}}
\put(11.70,9.20){\vector(1,0){6.00}}
\put(10.00,5.00){\vector(0,1){2.00}}
\put(10.00,3.00){\line(0,1){1.00}}
\put(10.00,7.80){\line(0,1){0.50}} \ \
\put(11.00,11.50){\line(1,0){0.80}}
\put(13.00,11.50){\line(1,0){0.80}}
\put(15.00,11.50){\vector(1,0){2.00}} \
\put(25.00,9.00){\vector(1,0){9.00}}
\put(38.00,9.00){\vector(1,0){9.00}}

\put(27.00,8.00){\line(0,1){1.00}}
\put(27.00,5.00){\vector(0,1){2.00}}
\put(27.00,3.00){\line(0,1){1.00}}

\put(32.00,9.00){\line(0,1){1.00}}
\put(32.00,11.00){\vector(0,1){2.00}}
\put(32.00,14.00){\line(0,1){1.00}}

\put(40.00,9.00){\line(0,1){1.00}}
\put(40.00,11.00){\vector(0,1){2.00}}
\put(40.00,14.00){\line(0,1){1.00}}

\put(45.00,8.00){\line(0,1){1.00}}
\put(45.00,5.00){\vector(0,1){2.00}}
\put(45.00,3.00){\line(0,1){1.00}}

\put(21.00,9.00){\makebox(0,0)[cc]{=}}
\put(36.00,9.00){\makebox(0,0)[cc]{+}}

\end{picture}
 \caption{Compton scattering of equivalent photon on the electron.}
 \label{fig:2}
\end{figure}
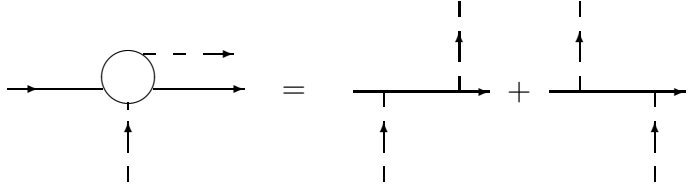

%%%%%%%%%%%%%%%%%%%%%%%%%%%%%%%%%%%%%%%%%%%%%%%
\section{Qualitative description of the MD-effect}
%%%%%%%%%%%%%%%%%%%%%%%%%%%%%%%%%%%%%%%%%%%%%%%%

Qualitatively we describe the MD--effect using the $e p
\rightarrow e p \gamma$ process as an example. This reaction is
described by the diagram of Fig.~\ref{fig:1} which corresponds to
the radiation of the photon by the electron (the contribution of
the photon radiation by the proton can be neglected). The large
impact parameters $\varrho \gtrsim \sigma_\perp$, where
$\sigma_\perp$ is the transverse beam size, correspond to small
momentum transfer $\hbar q_\perp \sim (\hbar / \varrho) \lesssim
(\hbar / \sigma_\perp)$. In this region, the given reaction can be
represented as a Compton scattering (Fig.~\ref{fig:2}) of the
equivalent photon, radiated by the proton, on the electron. The
equivalent photons with frequency $\omega$ form a ``disk'' of
radius $\varrho_m \sim \gamma_p c / \omega$ where $\gamma_p = E_p
/ (m_p c^2)$ is the Lorentz-factor of the proton. Indeed, the
electromagnetic field of the proton is $\gamma_p$--times
contracted in the direction of motion. Therefore, at distance
$\varrho$ from the axis of motion a characteristic longitudinal
length of a region occupied by the field can be estimated as
$\lambda \sim \varrho / \gamma_p$ which leads to the frequency
$\omega \sim c / \lambda \sim \gamma_p c / \varrho$.

\begin{figure}[!ht ]
  \centering
\unitlength=2.00mm \special{em:linewidth 0.4pt}
\linethickness{0.4pt}
\begin{picture}(47.00,15.00)

\put(13.00,7.00){\oval( 8,14)}

\put(13.00 ,6.85){\line(1,0){6.30}}
 \put(13.00,7.15){\line(1,0){6.30}}
 \put(13.00 ,6.90){\line(0,1){7.00}}
\put(19.40 ,7.00){\vector(1,0){0.50}}

\put(14.50 ,8.50){\line(1,0){0.70}}
 \put(15.70,8.50){\line(1,0){0.70}}
  \put(17.00,8.50){\line(1,0){0.70}}
 \put(18.20 ,8.50){\vector(1,0){2.00}}

\put(14.50 ,9.50){\line(1,0){0.70}}
\put(15.70,9.50){\line(1,0){0.70}}
 \put(17.00,9.50){\line(1,0){0.70}}
\put(18.20 ,9.50){\vector(1,0){2.00}}

\put(14.50 ,10.50){\line(1,0){0.70}}
 \put(15.70,10.50){\line(1,0){0.70}}
  \put(17.00,10.50){\line(1,0){0.70}}
\put(18.20 ,10.50){\vector(1,0){2.00}}

\put(14.50 ,5.50){\line(1,0){0.70}}
 \put(15.70,5.50){\line(1,0){0.70}}
  \put(17.00,5.50){\line(1,0){0.70}}
  \put(18.20 ,5.50){\vector(1,0){2.00}}

\put(14.50 ,4.50){\line(1,0){0.70}}
 \put(15.70,4.50){\line(1,0){0.70}}
  \put(17.00,4.50){\line(1,0){0.70}}
 \put(18.20 ,4.50){\vector(1,0){2.00}}

\put(14.50 ,3.50){\line(1,0){0.70}}
 \put(15.70,3.50){\line(1,0){0.70}}
  \put(17.00,3.50){\line(1,0){0.70}}
\put(18.20 ,3.50){\vector(1,0){2.00}}

\put(11.50,10.00){\makebox(0,0)[cc]{$\varrho_m$}}
\put(11.50,7.00){\makebox(0,0)[cc]{p}}
\put(19.00,2.00){\makebox(0,0)[cc]{$\omega$}}

\put(31.00,7.00){\oval( 10,4)}

\put(26.00,7.00){\vector(-1,0){2.00}}
\put(31.00,10.50){\vector(0,-1){1.50}}
\put(31.00,3.50){\vector(0,1){1.50}}

\put(31.00,7.00){\makebox(0,0)[cc]{$\sigma_\perp$}}
\put(25.00,5.00){\makebox(0,0)[cc]{e}}

\end{picture}
\caption{Scattering of equivalent photons, forming the ``disk"
with radius $\varrho_m$, on the electron beam with radius
$\sigma_\perp$. }
 \label{fig:3}
\end{figure}
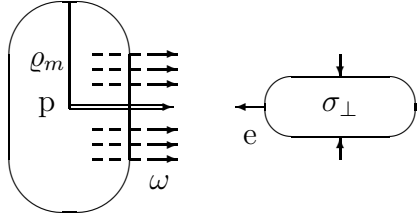

In the reference frame connected with the collider, the equivalent
photon with energy $\hbar \omega$ and the electron with energy
$E_e \gg \hbar \omega$ move toward each other (Fig.~\ref{fig:3})
and perform the Compton scattering. The characteristics of this
process are well known. The main contribution to the Compton
scattering is given by the region where the scattered photons fly
in a direction opposite to that of the initial photons. For such a
backward scattering, the energy of the equivalent photon $\hbar
\omega$, the energy of the final photon $E_\gamma$, and its
emission angle $\theta_\gamma$ are related by
 \begin{equation}
    \hbar \omega = {E_\gamma \over 4 \gamma^2_e (1 - E_\gamma/E_e )}
   \left[1+ (\gamma_e\theta_\gamma)^2 \right]
 \label{1.1a}
 \end{equation}
and, therefore, for the typical emission angles
$\theta_\gamma\lesssim 1/\gamma_e$ we have
\begin{equation}
 \hbar \omega \sim {E_\gamma \over 4 \gamma^2_e (1 - E_\gamma/E_e
   )}\,.
 \label{1.1}
 \end{equation}

As a result, we find the radius of the ``disk'' of equivalent
photons with the frequency $\omega$ (corresponding to a final
photon with energy $E_\gamma$) as follows:
 \begin{equation}
 \varrho_m = {\gamma_p c\over  \omega} \sim 4\, \lambda_e\; {\gamma_e
\gamma_p}\, {E_e - E_\gamma \over E_\gamma} \,,\;\;\;
\lambda_e={\hbar\over m_ec}=3.86\cdot 10^{-11}\; {\rm cm}\,.
 \label{size}
 \end{equation}
For the HERA collider with $E_p=820$ GeV and $E_e=28$ GeV one
obtains
 \begin{equation}
 \varrho_m \gtrsim 1\; {\rm cm \ \ \ for\ \ \ } E_\gamma \lesssim
0.2 \;{\rm GeV}\ .
 \label{1.3}
 \end{equation}
Equation (\ref{size}) is also valid for the $e^- e^+ \rightarrow
e^- e^+ \gamma$ process with replacement the protons by the
positrons. For the Super$B$ factory~\cite{CDR} it leads to
 \begin{equation}
 \varrho_m \gtrsim 1\; {\rm cm \ \ \ for\ \ \ } E_\gamma \lesssim
0.1\; \mbox{ GeV }\,.
  \label{1.4a}
  \end{equation}

The  standard  calculation corresponds to the interaction of the
photons (that form the ``disk'') with the unbounded flux of
electrons. However, the particle beams at the HERA collider have
finite transverse beam sizes of the order of $\sigma_\perp\sim
10^{-2}$ cm. Therefore, the equivalent photons from the region
$\sigma_\perp \lesssim \varrho \lesssim \varrho_m$ cannot interact
with the electrons from the other beam. This leads to the
reduction of the number of the observed photons. The ``observed
cross section''  $d \sigma_{\rm obs}$ is smaller than the standard
cross section $d \sigma$ calculated for an infinite transverse
extension of the electron beam,
 \begin{equation}
  d \sigma - d \sigma_{\rm obs} = d \sigma_{\rm cor}.
  \label{1.5}
  \end{equation}
Here the correction $d \sigma_{\rm cor}$ can be presented in the
form
 \begin{equation}
 d \sigma_{\rm cor} = d \sigma_{\rm C}(\omega,\,E_e,\,E_\gamma) \
dn(\omega)
  \label{1.6}
  \end{equation}
where $dn(\omega)$ denotes the number of ``missing'' equivalent
photons and $d \sigma_{\rm C}$ is the cross section of the Compton
scattering.  Let us stress that the equivalent photon
approximation in this region has a high accuracy (the neglected
terms are of the order of $1/\gamma_p$). But for the qualitative
description it is sufficient to use the logarithmic approximation
in which this number is (see\cite{BLP}, \S 99)
 \begin{equation}
  dn = {\alpha \over \pi} {d\omega \over \omega} {d q_{\perp}^2 \over
q_{\perp}^2} \,.
 \label{1.7}
 \end{equation}
Since $q_\perp  \sim 1 / \varrho$, we can present the number of
``missing'' equivalent photons in another form
 \begin{equation}
 dn = {\alpha \over \pi} {d \omega \over \omega} { d\varrho^2 \over
\varrho^2}
 \label{1.8}
 \end{equation}
with the integration region in $\varrho$:
 \begin{equation}
 \sigma_\perp \lesssim \varrho \lesssim \varrho_m = {\gamma_p c
\over \omega}\,.
 \label{1.9}
 \end{equation}
As a result, this number is equal to
 \begin{equation}
 dn(\omega) = 2 {\alpha \over \pi} { d\omega \over \omega} \ln
{\varrho_m \over \sigma_\perp }\,,
 \label{1.10}
 \end{equation}
and the correction to the standard cross section with logarithmic
accuracy is (more exact expression is given by Eq. (\ref{18}))
 \begin{equation}
  d\sigma_{\rm cor} = {16\over 3} \alpha r^2_e\, {dy\over y}\,
\left(1-y+\mbox{${3\over 4}$} y^2\right) \ln{4\gamma_e \gamma_p
(1-y)\lambda_e\over y \sigma_\perp}\,, \;\;y={E_\gamma\over
E_e}\,.
  \label{1.11}
 \end{equation}

%%%%%%%%%%%%%%%%%%%%%%%%%%%%%%%%%%%%%%%%%%%
\section{MD-effect for Super$B$ factory}
%%%%%%%%%%%%%%%%%%%%%%%%%%%%%%%%%%%%%%%%%%%%%

Usually in experiments the cross section is found as the ratio of
the number of observed events per second $d\dot N$ to the
luminosity $L$. Also, in our case it is convenient to introduce
the ``observed cross section'', defined as the ratio
 \begin{equation}
d\sigma_{{\rm obs}} ={d\dot N \over L}\,.
 \end{equation}
Contrary to the standard cross section $d\sigma$, the observed
cross section $ d\sigma_{{\rm obs}}$ depends on the parameters of
the colliding beams. To indicate explicitly this dependence we
introduce the ``correction cross section'' $d\sigma_{{\rm cor}}$
as the difference between $d\sigma$ and $ d\sigma_{{\rm obs}}$:
 \begin{equation}
d\sigma_{{\rm obs}} = d\sigma - d\sigma_{{\rm cor}}\,.
 \end{equation}
The relative magnitude of the MD-effect is given, therefore, by
quantity
 \begin{equation}
 \delta = { d\sigma_{{\rm cor}}\over d\sigma} \,.
 \label{delta}
 \end{equation}
Let us consider the number of photons emitted by electrons in the
process $e^-e^+ \to e^- e^+ \gamma$. The standard cross section
for this process is well known:
 \begin{equation}
d\sigma^{(e)} = {16\over 3} \alpha r^2_e\, {dy\over y}\,
\left(1-y+{3\over 4} y^2\right)\,\left[\ln{4\gamma_e \gamma_p
(1-y)\over y} \, -\, {1\over 2}\right]\,,\;\; y={E_\gamma\over
E_e}\,,
 \label{17}
 \end{equation}
where $\gamma_e=E_e/(m_e c^2)$ and $\gamma_p=E_p/(m_e c^2)$ is the
Lorentz-factor for the electron and positron, respectively,
$\alpha =e^2/(\hbar c) \approx 1/137$ and $r_e=e^2/(m_e c^2)$.

The correction cross section depends on the r.m.s. transverse
horizontal and transverse vertical bunch sizes $\sigma_{jH}$ and
$\sigma_{jV}$ for the electron, $j=e$, and positron, $j=p$, beams.
Besides, for the the considered collider we have to take into
account that its $e^{\pm}$ beams of the length
$l_e=l_p\equiv\sigma_z$ collide to a crossing angle $2\psi$. In
calculations below we used data from Conceptual Design
Report~\cite{CDR} (see Table~1).%\ref{tab1}).

 \vspace{3mm}

\begin{table}[htb]
  \label{tab1}
\begin{center}
\centerline{Table 1:Parameters of beams used for calculations}
 \vspace{3mm}
\par
\renewcommand{\arraystretch}{1.5}
\begin{tabular}{|c|c|c|c|c|c|c|c|c|c|} \hline
$E_e$, & $E_p$, & $\sigma_V$, &  $\sigma_H$, & $\sigma_z$ &
$2\psi$ & Energy
\\ %\hline
 GeV &  GeV & $\mu$m & $\mu$m & cm & mrad & spread, \%
\\ \hline
 7 & 4 & 0.035 & 5.657 & 0.6 & $34$ & 0.09
 \\ \hline
\end{tabular}
\end{center}
 \end{table}
Formulas of the correction cross section for this case have been
obtained in~\cite{KPS85b}. In the above notations the correction
cross section is as follows:
 \begin{equation}
d\sigma^{(e)}_{\rm cor} = {16\over 3} \alpha r^2_e\, {dy\over y}\,
\left[\left(1-y+{3\over 4} y^2\right) L_{\rm cor}
 -{1-y\over 12}\right]
 \label{18}
 \end{equation}
where
 \begin{eqnarray}
L_{\rm cor}&=& \ln{2\sqrt{2}\gamma_e \gamma_p
(1-y)(a_H+a_V)\lambda_e \over  a_H a_V y} \, -\, {3+C\over
 2}\,,
 \nonumber\\
\lambda_e&=&{\hbar\over m_ec}=3.86\cdot 10^{-11}\; {\rm cm}\,,
\;\;\; C=0.577...\,, \\ a_H&=&\sqrt{\sigma_{eH}^2+
\sigma_{pH}^2+(l_e^2+l_p^2)\, \psi^2}\,, \;\;
a_V=\sqrt{\sigma_{eV}^2+ \sigma_{pV}^2}\,.
 \nonumber
 \end{eqnarray}

\begin{figure}[!ht]
 \centering
\vspace{5mm}
\begin{picture}(340,160)
 \put(230,-5){\makebox(0,0)[t]{$y=E_\gamma/E_e$}}
 \put(245,145){\makebox(0,0)[t]{$\fr{y}{\alpha r_e^2}\,\fr{d\sigma^{(e)}}{dy}$}}
 \put(55,115){\makebox(0,0)[t]{$\fr{y}{\alpha r_e^2}\,\fr{d\sigma^{(e)}_{\rm obs}}{dy}$}}
\includegraphics[width=0.7\textwidth]{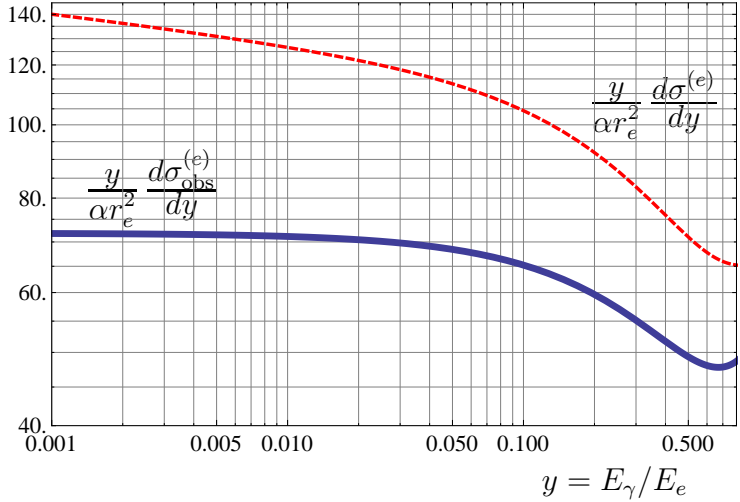}
\end{picture}
 \vspace{5mm}
  \caption{The standard cross section $(y/\alpha r_e^2)\,(d\sigma^{(e)}/dy)$ (the
dashed curve) and the cross section with the beam-size correction
$(y/\alpha r_e^2)\,(d\sigma^{(e)}_{\rm obs}/dy)$ (the solid curve)
versus the relative photon energy $y=E_\gamma /E_e$ for the
Super$B$ factory}
 \label{fig:4}
  \end{figure}

The observed number of photons is smaller due to MD-effect than
the number of photons calculated without this effect (Fig.
\ref{fig:4}).

The relative magnitude of the MD-effect is given by quantity
$\delta$ from Eq. (\ref{delta}) (see Table~2). It can be seen from
Fig.~\ref{fig:4} and Table~2 that the MD-effect considerably
reduces the differential cross section.

 \vspace{3mm}

\begin{table}[htb]
%\label{tab2}
 \centerline{Table 2: Relative magnitude of the MD-effect for different photon
energies}
 \vspace{3mm}
\begin{center}
\par
\renewcommand{\arraystretch}{1.5}
\begin{tabular}{|c|c|c|c|c|c|c|} \hline
$y=E_\gamma/E_e$ & 0.001 & 0.005 & 0.01 & 0.05 & 0.1 & 0.5
\\ \hline
$\delta,\; \%$  & 49 & 45 & 44 & 40 & 38 & 31 \\ \hline
\end{tabular}
\end{center}
 \end{table}

To estimate the integrated contribution of the discussed process
into particle losses, we should integrate the differential
observed cross section from some minimal photon energy. It is
usually assumed that an electron leaves the bunch when it emits
the photon with the energy either larger than $1$ \% of the
electron energy or $10$ times larger than the beam energy spread.
In other words, the relative photon energy $y=E_\gamma/E_e$ should
be larger than $y_{\min}=0.01$ or $y_{\min}= 0.009$. In
calculations below we use $y_{\min}=0.01$. After integration of
the differential observed cross section from $y_{\min}\ll 1$ up to
$y_{\max}=1$, we obtain
\begin{eqnarray}
\sigma_{\rm obs}^{(e)}& =& {16\over 3} \alpha r^2_e\,\left\{
 \left(\ln{1\over y_{\min}}- {5\over 8}\right)
\left[ \ln{\sqrt{2} a_H a_V \over (a_H+a_V)\lambda_e} \, +\,
{2+C\over
 2} \right] \right.
 \nonumber\\
 &+& \left.
 {1\over 12}\left(\ln{1\over y_{\min}}- 1 \right)
 \right\}
 \end{eqnarray}
or
\begin{equation}
\sigma_{\rm obs}^{(e)}(y_{\min}= 0.01)=166 \;\mbox{mbarn}\,.
 \label{ourres}
 \end{equation}

Let us note that the standard cross section integrated over the
same interval of $y$, is equal to
 \begin{eqnarray}
\sigma^{(e)} &=&{16\over 3} \alpha r^2_e\,\left\{\left(\ln{1\over
y_{\min}}- {5\over 8}\right) \,\left[\ln{(4\gamma_e \gamma_p)} \,
-\, {1\over 2}\right] \right.
 \nonumber\\
 &+& \left.
{1\over 2}\, \left(\ln{1\over y_{\min}}\right)^2  -{3\over
8}-{\pi^2\over 6}\right\}
 \end{eqnarray}
or
\begin{equation}
\sigma^{(e)}(y_{\min}= 0.01)=265 \;\mbox{mbarn}\,.
 \end{equation}

Therefore, the observed cross section is smaller than the standard
one by $37$~\%.

\section{Discussion}

In conclusion, we have calculated the MD-effect at the Super$B$
factory. We find out that this effect reduces beam particle losses
due to bremsstrahlung by about 40\%.

Then we compare our result (\ref{ourres}) with that presented in
Sec. 3.6.2 of the paper~\cite{CDR}:
\begin{equation}
\sigma_{\rm obs}^{(e){\rm CDR}}(y_{\min}= 0.01)=170
\;\mbox{mbarn}\,.
 \label{CDRres}
 \end{equation}
We found out the good agreement between these two results.
Unfortunately, this agreement is nothing else but a simple {\bf
random coincidence} because {\bf the base of our approach and
approach used in~\cite{CDR} is quite different}.

In a few words, the essence of our approach is the following.
Those impact parameters $\varrho$, which give an essential
contribution to the standard cross section at the discussed
collider, reach values of $\varrho_m \sim 2$ cm at $E_\gamma =
0.01\,E_e$ whereas the transverse size of the bunch is of the
order of transverse bunch size $\sigma_V$. The limitation of the
impact parameters to values
 \be
\varrho \lesssim \sigma_V=0.035\;\mu\mbox{m}
 \ee
is just the reason for the decreasing number of observed photons.

On the other hand, the results in CDR is based on BBBREM Monte
simulation code of Ref.~\cite{KB-94a} which used a formula for the
distance cut-off given in Ref.~\cite{BK-94b}. Authors of
Refs.~\cite{KB-94a},~\cite{BK-94b} call this phenomenon as a
density effect and used the cut-off at half the average distance
$d$ between two positrons in the bunch at rest (if we speak about
emission by electrons). It correspond the limitation of the impact
parameters to values
 \be
\varrho \lesssim d=\frac 12 \left(\frac{\sigma_V \sigma_H \gamma_p
\sigma_z}{N_p}\right)^{1/3} =0.032\;\mu\mbox{m}\,.
  \ee
The random coincidence of these two values is the origin of a good
agreement between two results (\ref{ourres})  and (\ref{CDRres}) .
However, we should know which approach is correct, in order to
understand tendencies in the case when some parameters of the
collider will change.

We have a strong doubt about approach used in
Refs.~\cite{KB-94a},~\cite{BK-94b}. From the theoretical point of
view, we do not see any clear explanation, but a simple recipe.
Besides, it contradicts to the existing HERA experiment
\cite{Piot95}.

On the contrary, our approach has a clear qualitative explanation
given in Sec.~2. Our calculations are based on such a solid theory
as QED and are confirmed by a number of experiments at the VEPP-4
collider in BINP (Novosibirsk) and at the HERA collider in DESY.
In particular, in the VEPP-4 experiments \cite{Blinov82},
\cite{Tikhonov82} and \cite{Blinov88} it was studied not one but
several different quantities, including the measurement of the
effective cross section as function of the transverse beam
parameters (from $\sigma_V=10\;\mu$m to $\sigma_V=60\;\mu$m) and
dependence of the photon rate on the shift of one bunch in the
vertical direction on the distance up to $3 \sigma_V$. All these
measurements supported the concept that the effect arises from the
limitation of the impact parameters.

Certainly, accuracy of all experiments is far from excellent and
further investigations are desirable, but from experimental point
of view just now there is no another explanation with such a solid
base as the MD-explanation.

\section*{Acknowledgments}

We are very grateful to A.~Bondar,  I.~Koop and A.~Onuchin for
useful discussions. This work is partially supported by the
Russian Foundation for Basic Research (code 09-02-00263) and by
the Fond of Russian Scientific Schools (code 1027.2008.2).

\end{document}